\documentclass[a4paper,12pt]{article}

\usepackage[latin1]{inputenc} 
\usepackage[T1]{fontenc}

\usepackage{dsfont}
\usepackage{amstext}  
\usepackage{amsfonts}
\usepackage{amssymb} 
\usepackage{amsmath}
\usepackage{amsthm}      
\usepackage{xcolor} 
\usepackage{color}
\usepackage{array}                
\usepackage{framed}

\usepackage{dsfont}
\usepackage{multirow}
\usepackage{graphicx}
\usepackage{axodraw4j}

\newcommand{\SU}{\ensuremath{\mathrm{SU}}}
\newcommand{\p}{\partial}

\newcommand{\sss}[1]{\scriptscriptstyle#1}
\newcommand{\ssst}[1]{\scriptscriptstyle\text{#1}}
\newcommand{\vv}[2]{\left( \begin{array}{c} #1 \\ #2  \end{array} \right)}
\newcommand{\bea}{\begin{eqnarray}}
\newcommand{\eea}{\end{eqnarray}}
\newcommand{\be}{\begin{equation}}
\newcommand{\ee}{\end{equation}}
\newcommand{\ba}{\begin{align}}
\newcommand{\ea}{\end{align}}
\newcommand{\beas}{\begin{eqnarray*}}
\newcommand{\eeas}{\end{eqnarray*}}
\newcommand{\bes}{\begin{equation*}}
\newcommand{\ees}{\end{equation*}}
\newcommand{\bas}{\begin{align*}}
\newcommand{\eas}{\end{align*}}
 \newcommand{\rig}{\rightarrow}
 
\newcommand{\ssL}{{\mathcal L}}

\newcommand{\gs}{g_{\scriptscriptstyle{s}}}
\newcommand{\yt}{y_{\scriptscriptstyle{t}}}

\newcommand{\gb}{g_1}
\newcommand{\gw}{g_2}

\newcommand{\als}{\alpha_{\scriptscriptstyle{s}}}

\newcommand{\lb}{\left(}
\newcommand{\rb}{\right)}

\newcommand{\msbar}{$\overline{\text{MS}}$}

\definecolor{bluemar}{rgb}{0,0,.5}
\definecolor{redmar}{rgb}{.8,0,0}
\definecolor{greenmar}{rgb}{0,.5,0}

\begin{document}

\begin{flushright}
TTP14-030,\\ SFB/CPP-14-87
\par\end{flushright}

\begin{center}
{\huge\bf Standard Model beta-functions to\\[1mm] three-loop order and vacuum stability}
\vskip 0.3cm
 Max F. Zoller$^{\;a,1}$,\\[1ex]
{\small $^{a}$ Institut f\"ur Theoretische Teilchenphysik (TTP), Karlsruhe Institute for Technology}
\vskip 0.2cm
\end{center}
{\small $1$: max.zoller@kit.edu}

\vskip 0.2cm
\begin{center}
{\bf Abstract}\\[2ex]
\end{center}

%
%

\date{\today}

Since the discovery of a Higgs particle \cite{ATLAS:2012ae,Chatrchyan:2012tx} the effective Higgs potential
of the Standard Model or extensions and the stability of the ground state corresponding to its minimum at the electroweak scale have been
subject to a lot of investigation. The vacuum expectation value of the scalar $\SU(2)$ doublet 
field in the Standard Model, which is responsible for the masses of elementary particles, may in fact not be at the global 
minimum of the effective Higgs potential.
The question whether there is a deeper minimum at some large scale is closely linked to the behaviour of the running quartic
Higgs self-interaction $\lambda(\mu)$.
In this talk an update on the analysis of the evolution of this coupling is given. We use three-loop beta-functions
for the Standard Model couplings, two-loop matching between on-shell and \text{\msbar } quantities and compare the theoretical
precision achieved in this way to the precision in the latest experimental values for the key parameters.



\section{Introduction: The stability of the Standard Model ground state}
\begin{figure}[!ht]
\begin{tabular}{ccc}
\begin{picture}(130,90)(0,0)
\put(0,0){\includegraphics[width=0.31\linewidth]{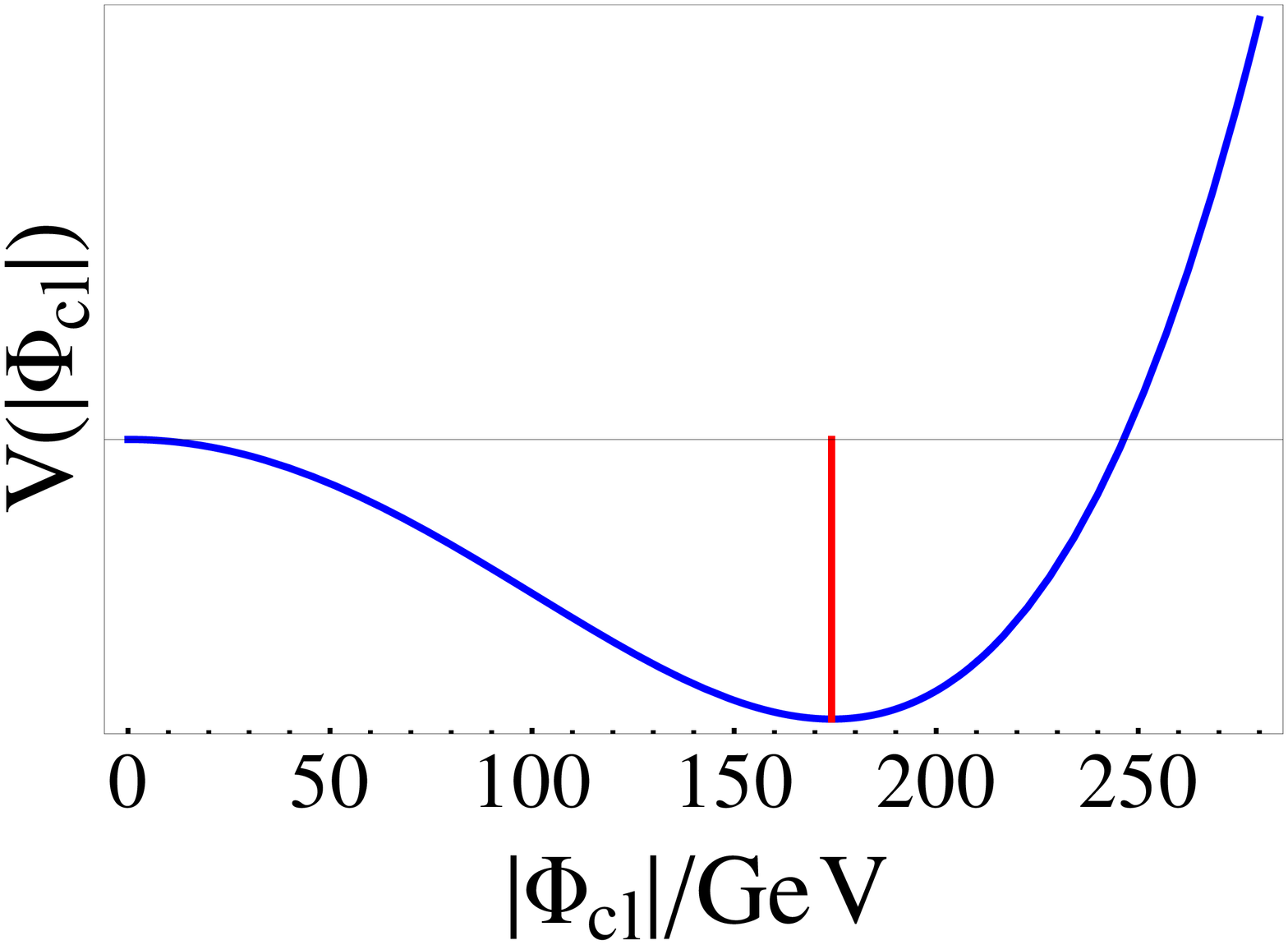}}
\Text(79,57)[cc]{{\Red{$\frac{v}{\sqrt{2}}$}}}
 \end{picture}
&
\begin{picture}(135,90)(0,0)
\put(0,0){\includegraphics[width=0.345\linewidth]{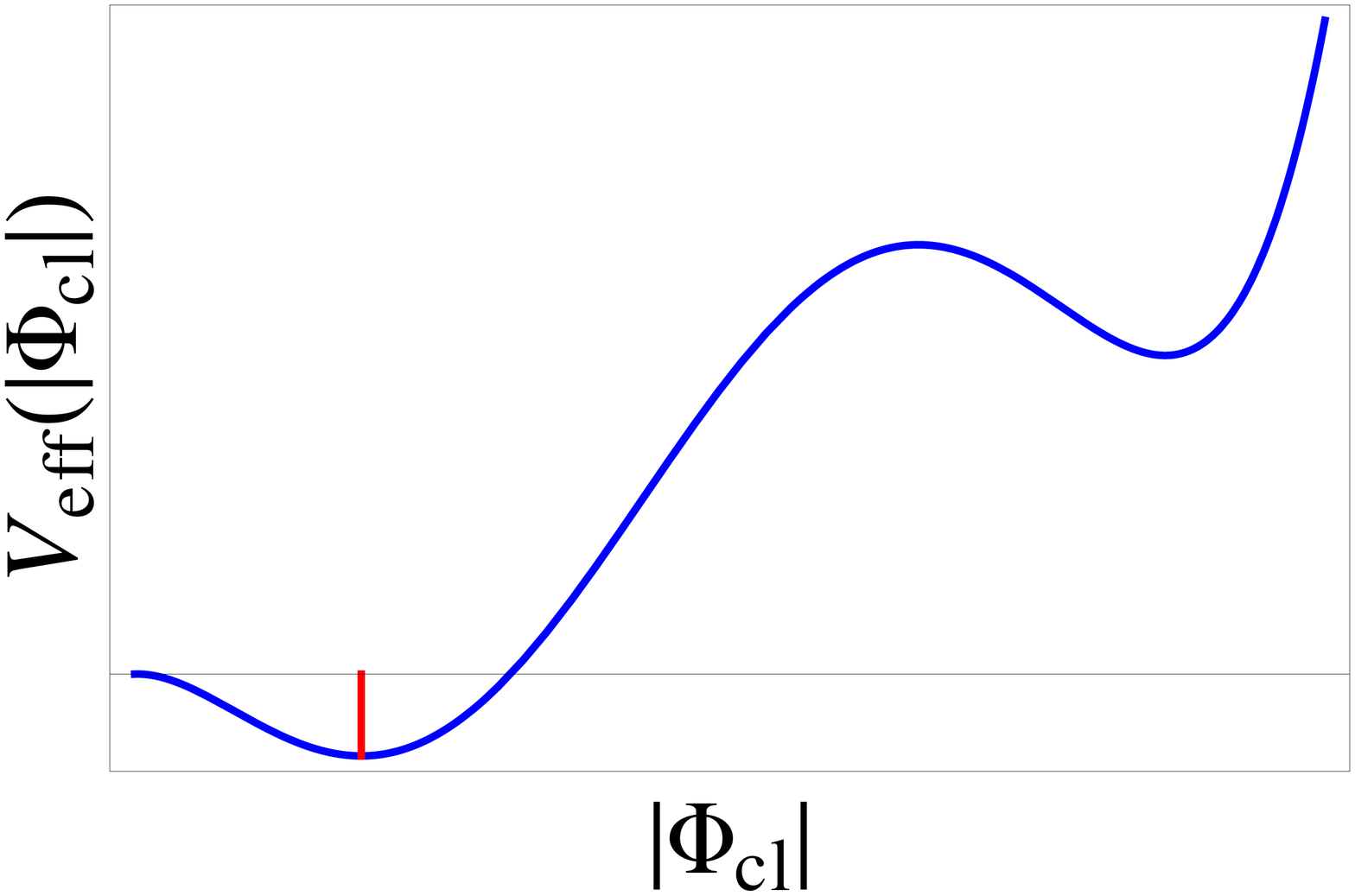}}
\Text(35,30)[cc]{{\Red{$\frac{v}{\sqrt{2}}$}}}
 \end{picture}
&
\begin{picture}(130,90)(0,0)
\put(0,0){\includegraphics[width=0.345\linewidth]{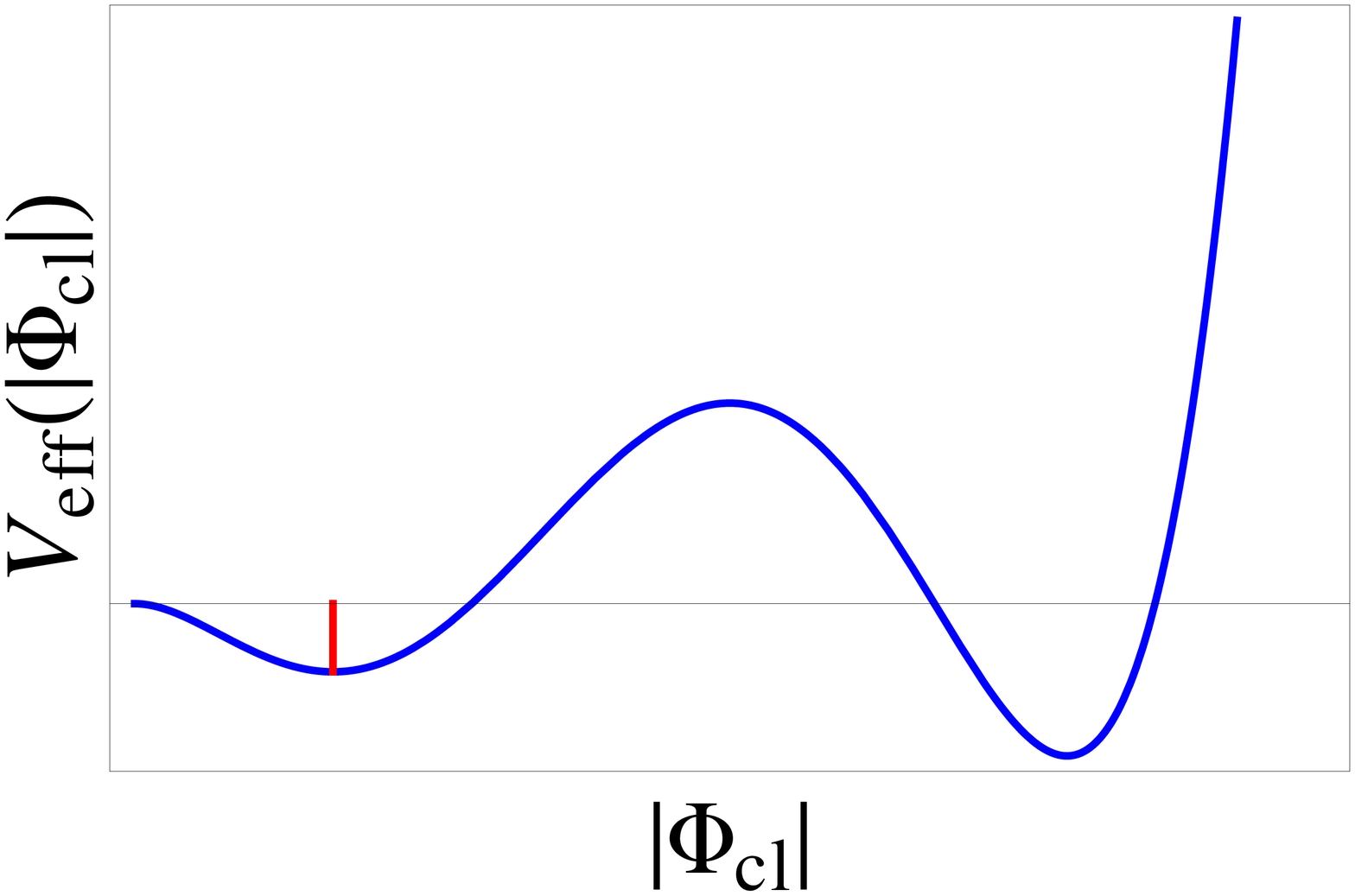}}
\Text(34,38)[cc]{{\Red{$\frac{v}{\sqrt{2}}$}}}
 \end{picture}\\[2ex]
%
%
(a) & (b) & (c)\\ \fbox{tree level} &
\fbox{$\lambda(\mu)>0\; \forall\; \mu\leq\Lambda$} &
\fbox{$\exists\; \mu\leq\Lambda:\; \lambda(\mu)<0$} 
\end{tabular}
\caption{Classical and effective Higgs potential as a function of $|\Phi_{\ssst{cl}}|:=\sqrt{\Phi^\dagger\Phi}$.} \label{Veffective} 
\end{figure}

In the Standard Model (SM) of particle physics fermions interact via the exchange of gauge bosons.
The strength of these interactions is given by the coupling constants $\gs$ for the QCD part and $\gw, \gb$ for the 
electroweak part. Furthermore, a scalar $\SU(2)$ doublet $\Phi=\vv{\Phi_1}{\Phi_2}$ is introduced
which couples to the $\SU(2)$ gauge bosons via the coupling $\gw$ and to the fermions via Yukawa couplings, 
the top-Yukawa coupling $\yt$ being the strongest.
The quartic self-coupling $\lambda$ of the field $\Phi$ appears in the classical Higgs potential
\be V(\Phi)=m^2\Phi^\dagger\Phi + \lambda \lb\Phi^\dagger\Phi\rb^2 \label{Higgspotentialklassisch}{}. \ee
For $m^2<0$ the doublet $\Phi$ aquires a vacuum expectation value (VEV) $\langle \Phi \rangle=\frac{1}{\sqrt{2}}\vv{0}{v}$ in the minimum of
the classical Higgs potential (Fig.~\ref{Veffective} (a)). The masses of the quarks, leptons, massive gauge bosons 
and the Higgs boson are then proportional to the value $v\approx 246.2$ GeV \cite{pdg2014}. 

If we assume the SM to be valid up to the Planck scale
$\Lambda\sim 10^{19}$ GeV -- a reasonable scenario in the absence of new physics -- we have to include quantum corrections
which change the shape of the effective Higgs potential \cite{PhysRevD.7.1888} significantly as compared to the 
classical potential at high scales. 

The effective potential is best introduced in the path integral approach to quantum field theory.
Consider the generating functional $W[J]$ defined by the path integral
\be
\text{e}^{i\,W[J]}:=\int\!\mathrm{d}\Phi\mathrm{d}A\,\text{e}^{i\int\!\mathrm{d}^4x 
\left[ \ssL(\Phi(x),\p_\mu \Phi(x),A(x),\p_\mu A(x))+J(x)\Phi(x)\right]}=\langle 0^+|0^-\rangle|_J{} \label{ZJdefinition}
\ee
which describes the transition from the vacuum state at $t\rig-\infty$ to the one at $t\rig+\infty$ in the presence of the external current
$J=\vv{J_1}{J_2}$ coupling to the scalar doublet $\Phi$. All other fields in the Lagrangian $\ssL$ are denoted by $A$.
We can eliminate the external current $J$ by a Legendre transformation introducing the effective action 
\be
\Gamma[\Phi_{\ssst{cl}}]:=W[J]-\int\!\mathrm{d}^4x J(x)\Phi_{\ssst{cl}}(x) \label{effactiondefinition}
\ee
and the classical field strength
\be 
\Phi_{\ssst{cl}}(x):=\frac{\delta W[J]}{\delta J(x)}=\left.\frac{\langle 0^+|\Phi(x)|0^-\rangle}{\langle 0^+|0^-\rangle}\right|_{J}. \label{phicldefinition}
\ee
The effective Higgs potential $V_{\ssst{eff}}$ is a function of $\Phi_{\ssst{cl}}$ and can be defined as the first term in an
expansion of the effective action around the point where all fields have zero momentum:
\be
\Gamma[\Phi_{\ssst{cl}}]=\int\!\mathrm{d}^4x\,
\lb -V_{\ssst{eff}}(\Phi_{\ssst{cl}})+\frac{1}{2} (\p_\mu\Phi_{\ssst{cl}})^2Z(\Phi_{\ssst{cl}})+\ldots\rb {}.\label{Gammaeffexpansion2}
\ee
The so-defined effective potential, which in general depends on all parameters of the theory, contains two main pieces of information. On the one hand the $n$th derivative wrt $\Phi_{\ssst{cl}}$ gives the
effective strength of the interaction of $n$ external scalar fields, e.g.
\be
\frac{\mathrm{d}^2V_{\ssst{eff}}}{\mathrm{d}\Phi_{\ssst{cl}}^2}=m_{\ssst{eff}}^2,
\qquad
\frac{\mathrm{d}^4V_{\ssst{eff}}}{\mathrm{d}\Phi_{\ssst{cl}}^4}=\lambda_{\ssst{eff}}{}.
\ee
On the other hand the requirement 
\be
\frac{\mathrm{d}V_{\ssst{eff}}}{\mathrm{d}\Phi_{\ssst{cl}}}=0{}. \label{BedingungVEVausVeff}
\ee
yields all the candidates for VEVs of the scalar field for $J=0$, which correspond to local minima in the effective potential.

Generic shapes of this effective potential are shown in Fig. \ref{Veffective} for the cases
of a Higgs mass larger (b) and smaller (c) than a critical value $m_{min}$, the minimal stability bound (see also \cite{Bezrukov:2012sa}).
If the second minimum at large scales is deeper than the first at the electroweak scale the latter is not stable against tunneling
to this global minimum\footnote{Note that the effective potential is a gauge dependent quantity (as are the renormalized field $\Phi$ and its
anomalous dimension $\gamma_{\sss{\Phi}}$) and hence
the exact location of the second minimum is also gauge dependent. The existence of a second minimum and the fact whether it is lower or higher
then the first, however, does not depent on the gauge parameters. For a recent discussion of this topic see \cite{DiLuzio:2014bua}.}.

For large field strengths \mbox{$\Phi_{\ssst{cl}}\sim\Lambda\gg v$} we can use the approximation \cite{Altarelli1994141}
\be V_{\ssst{eff}}(\Phi_{\ssst{cl}})\approx\lambda(\Phi_{\ssst{cl}}) \;\Phi_{\ssst{cl}}^4\; 
\lb \text{e}^{-\frac{1}{2}\int\limits_0^t\!\mathrm{d}t'\gamma_{\sss{\Phi}}(t')} \rb^4
\label{VeffRGdef}
\ee
with $t:=\ln\lb\frac{\Phi_{\ssst{cl}}^2}{v^2}\rb$ and the running coupling $\lambda(\mu)$ evolved to the
scale $\mu=\Phi_{\ssst{cl}}$. From this it has been demonstrated that the stability of the SM vacuum is in good approximation
equivalent to the question whether the running coupling $\lambda(\mu)$ stays positive up to the scale $\Lambda$
\cite{Altarelli1994141,Cabibbo:1979ay,Ford:1992mv}. It is this requirement which will be investigated at high precision
in this talk.

The vacuum stability problem has been subject to a lot of investigation over the last years
\cite{Zoller:2014xoa,Zoller:2013mra,Zoller:2012cv,Chetyrkin:2012rz,Buttazzo:2013uya,Masina:2012tz,Bezrukov:2012sa,Degrassi:2012ry,EliasMiro:2011aa,Holthausen:2011aa}.
For a recent discussion of the vacuum stability problem in the MSSM see \cite{Bobrowski:2014dla}.

\section{Calculations: $\beta$-functions and matching relations}
The evolution of the Higgs self-coupling $\lambda$ with the energy scale $\mu$ is given by the $\beta$-function
\be \beta_{\lambda}(\lambda,\yt,\gs,g_2,g_1,\ldots)=\mu^2\frac{d}{d\mu^2} \lambda(\mu). \label{betahiggs} \ee
This power series in the couplings of the SM is computed in perturbation theory and is available up to three-loop order
\cite{Chetyrkin:2012rz,Chetyrkin:2013wya,Bednyakov:2013eba,Bednyakov:2013cpa}
as well as the $\beta$-functions for the gauge \cite{PhysRevLett.108.151602,Mihaila:2012pz,Bednyakov:2012rb}
and Yukawa \cite{Chetyrkin:2012rz,Bednyakov:2012en} couplings, which are also needed in order to solve eq.~\eqref{betahiggs}
numerically. 
The second ingredient to the solution of eq.~\eqref{betahiggs} is a set of initial conditions
for each coupling, e.g. their values at the scale of the top mass $M_{\ssst{t}}$. These values are needed in the 
$\overline{\text{MS}}$-scheme in which the $\beta$-functions are computed. Matching relations 
between experimentally accessible on-shell quantities, such as the top quark pole mass $M_{\sss{t}}$ and Higgs pole mass $M_{\ssst{H}}$,
and $\overline{\text{MS}}$ parameters have been calculated at two-loop level \cite{Buttazzo:2013uya,Jegerlehner:2012kn,Bezrukov:2012sa,
Espinosa:2007qp,Hempfling:1994ar,Sirlin1986389}.
For the key parameters 
\bea
M_{\ssst{t}} &=& \lb 173.34 \pm 0.76\rb \text{ GeV \cite{ATLAS:2014wva}},\\
M_{\ssst{H}} &=& \lb 125.7\pm 0.4 \rb \text{ GeV \cite{pdg2014}},\\
\als^{\overline{\text{MS}}}(M_Z) &=& 0.1185 \pm 0.0006 \;\cite{pdg2014}
\eea
we find the following best values for the \text{\msbar} parameters:
\begin{table}[!h] \begin{center}
 \begin{tabular}{|l|l|}
  \hline
  $\gs(M_{\ssst{t}})$ & $1.1671$\\
  $g_2(M_{\ssst{t}})$ & $0.6483$\\
  $g_1(M_{\ssst{t}})$ & $0.3587$\\
  $\yt(M_{\ssst{t}})$ & $0.9369\pm 0.00050_{\text{(th,match)}}$\\
  $\lambda(M_{\ssst{t}})$ & $0.1272\pm 0.00030_{\text{(th,match)}}$\\
    \hline
 \end{tabular} \end{center} \label{initialcond} \caption{SM couplings in the $\overline{\text{MS}}$-scheme at $\mu=M_{\ssst{t}}$, 
theoretical uncertainties for $\yt$ and $\lambda$ stem from the on-shell to \text{\msbar} matching \cite{Buttazzo:2013uya}.}
\end{table} \newline
\section{Analysis: The evolution of $\lambda(\mu)$}
Applying three-loop $\beta$-functions for $\lambda,\yt,\gs,g_2$ and $g_1$ as well as the initial conditions from Tab.~\ref{initialcond}
we evaluate $\lambda(\mu)$ numerically up to \mbox{$\mu=\Lambda\sim10^{19}$ GeV}. 

\begin{figure}[!ht]
 \includegraphics[width=0.95\linewidth]{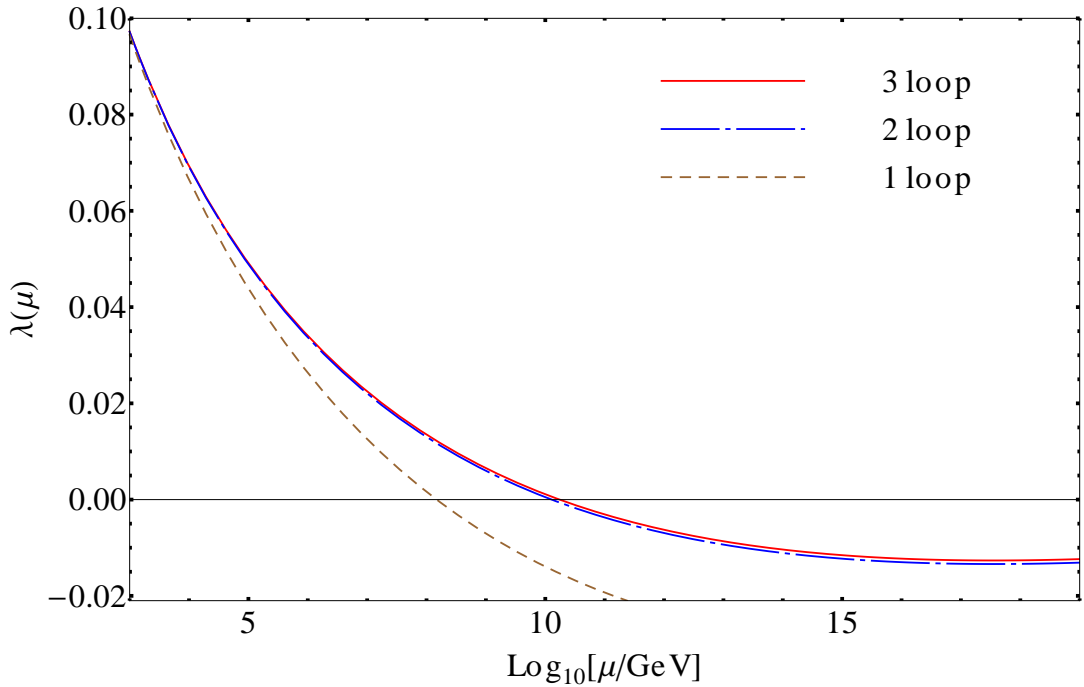}
\caption{Evolution of $\lambda$: One-, two- and three-loop beta-functions} \label{lambda_evolution_123} 
\end{figure}

Fig.~\ref{lambda_evolution_123} shows the results for one-loop, two-loop and three-loop $\beta$-functions which 
demonstrates the excellent convergence of the perturbation series. The difference between the two and three-loop curve can be taken
as an estimate for the theoretical uncertainty stemming from the $\beta$-functions.
From this we see that the SM ground state is no longer stable if we extend it to scales $\gtrsim 10^{10}$ GeV. This could be mended by some new physics
appearing between this scale and the electroweak scale. On the other hand -- as $\lambda$ stays close to zero -- the two minima of the effective
Higgs potential are almost degenerate in energy which leads to a lifetime of the electroweak ground state much longer than the age of the universe
\cite{Buttazzo:2013uya,Bezrukov:2012sa,Degrassi:2012ry,EliasMiro:2011aa}, and such a metastable scenario does not contradict our observations.

\begin{figure}[!ht]
 \includegraphics[width=0.95\linewidth]{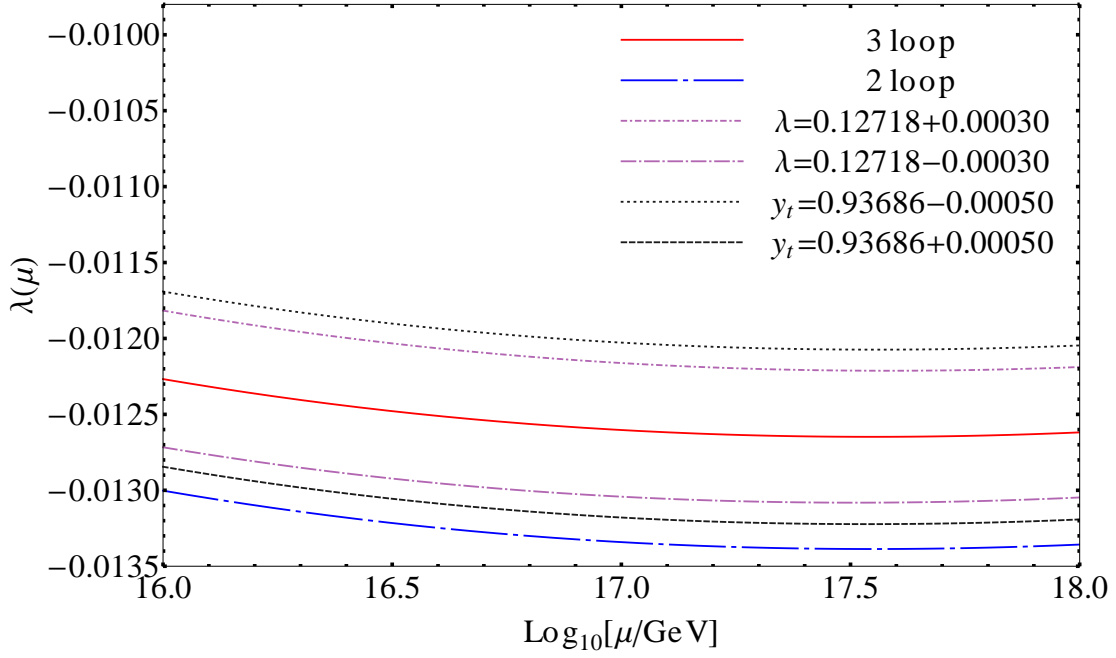}
\caption{Evolution of $\lambda$: Matching uncertainties} \label{lambda_evolution_Match} 
\end{figure}

But in order to give a definitive answer to the question whether the SM is stable up to large scales we have to consider all sources
of uncertainty. Apart from the small uncertainty stemming from the $\beta$-functions there is also a matching uncertainty of which the two main
contributions, the initial value for $\lambda$ and for $\yt$, are given in Tab.~\ref{initialcond}. The effect of varying these two parameters
by one $\sigma_{\ssst{matching}}$ for the three-loop curve is shown in Fig.~\ref{lambda_evolution_Match}. From this we can estimate that the matching precision 
is comparable to the precision in the $\beta$-functions.

\begin{figure}[!ht]
 \includegraphics[width=0.95\linewidth]{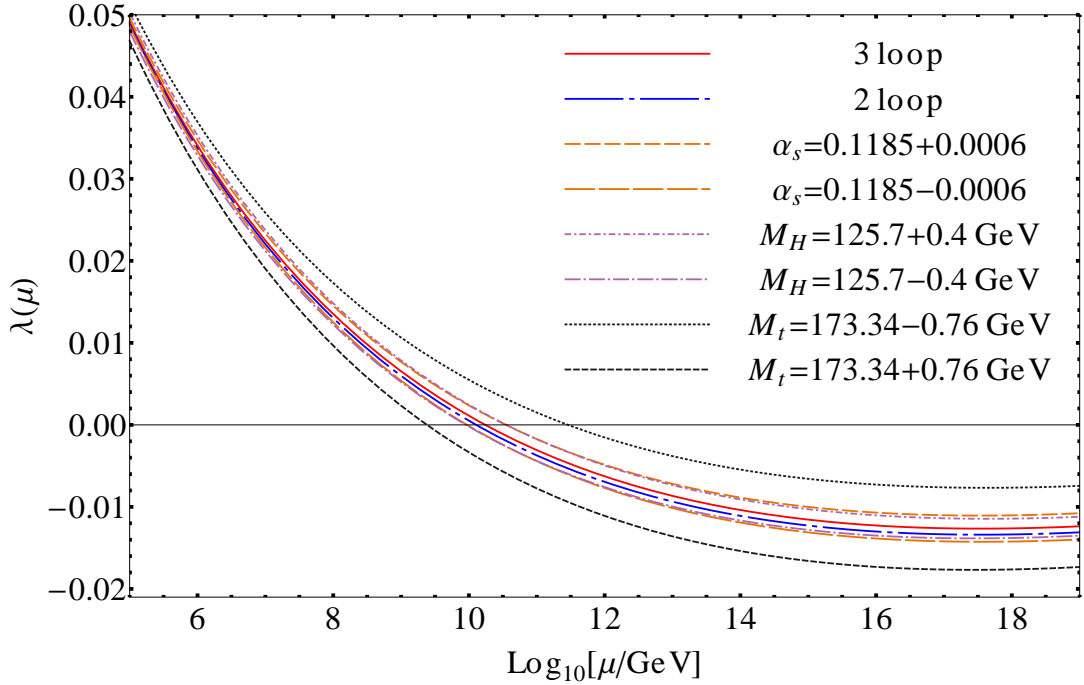}
\caption{Evolution of $\lambda$: Experimental uncertainties} \label{lambda_evolution_Exp} 
\end{figure}

By comparison the experimental uncertainties are significantly larger. The dashed (dotted) lines in Fig.~\ref{lambda_evolution_Exp} 
show the behaviour of $\lambda$ evolved using three-loop $\beta$-functions but with $M_{\ssst{t}}$, $M_{\ssst{H}}$ and $\als$ increased (decreased) 
by one standard deviation.

\begin{figure}[!ht]
 \includegraphics[width=0.95\linewidth]{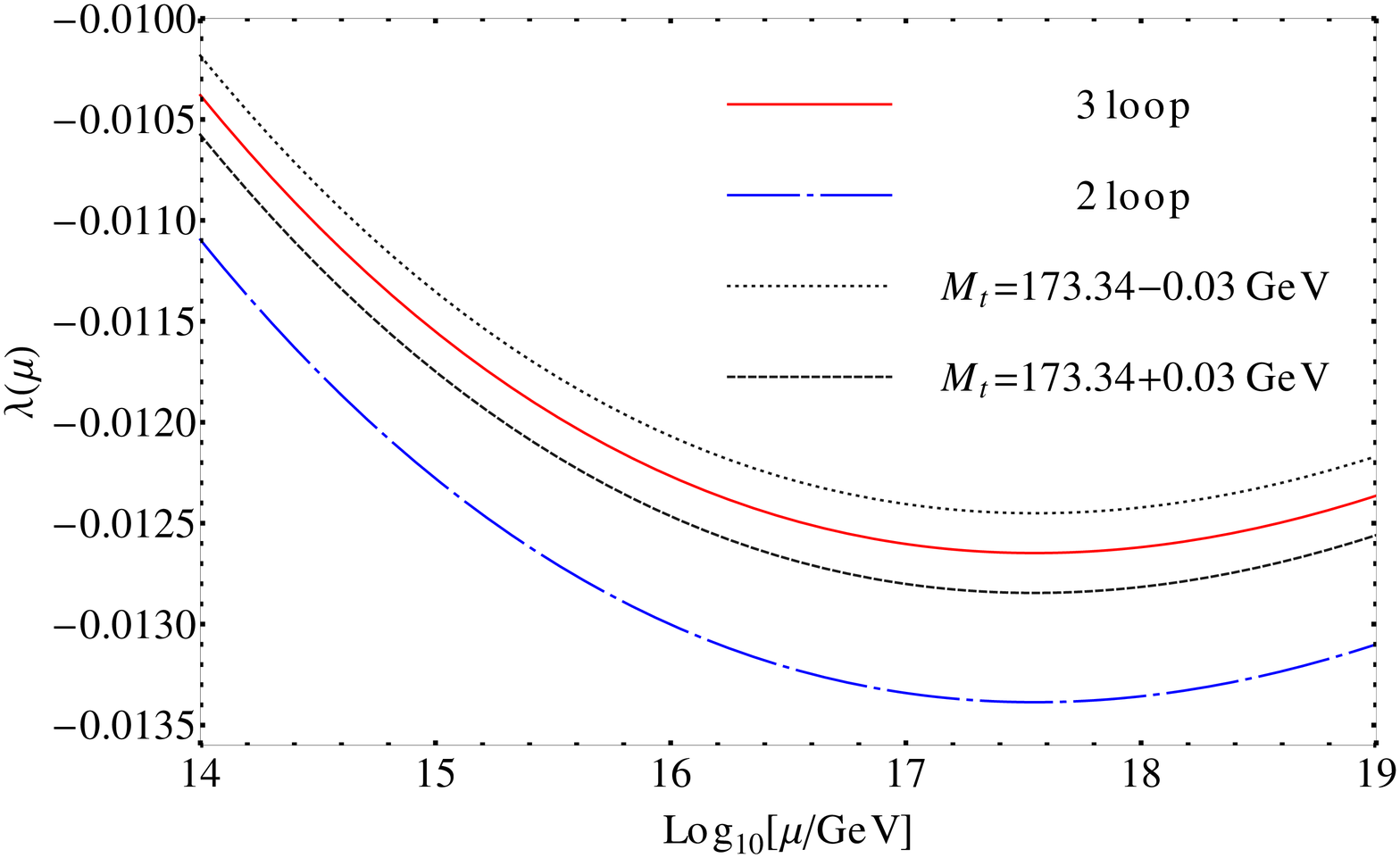}
\caption{Evolution of $\lambda$: Top mass uncertainty at the ILC} \label{lambda_evolution_Mt_ILC} 
\end{figure}

While the uncertainties originating from $\als$ and $M_{\ssst{H}}$ are approximately of the same size and at the Planck scale 
about a factor $2$ larger than the difference between the two-loop and three-loop curves, the uncertainty stemming from the top mass
measurement is about an order of magnitude larger than the theoretical one at $\mu\sim 10^{19}$ GeV. Within one $\sigma$ we are clearly in the 
metastable scenario for the SM but the large uncertainty in the top mass measurement does not allow for a final answer to the question of
vacuum stability. It is interesting to investigate the possibilities offered by a precision determination of the top mass, e.g.
at the ILC, where an uncertainty of \mbox{$\sigma_{M_{\ssst{t}}}\sim 30$ MeV} is within reach \cite{Martinez:2002st}.
Similarly, an uncertainty \mbox{$\sigma_{M_{\ssst{t}}}< 100$ MeV} is anticipated for CLIC \cite{Seidel:2013sqa}. In Fig.~\ref{lambda_evolution_Mt_ILC}
the evolution of $\lambda$ is shown for values of the top mass varied by the prospective \mbox{$\sigma_{M_{\ssst{t}}}= 30$ MeV}
which leads to an experimental uncertainty for this parameter which would be competitive with the theory uncertainty
for the evolution of the Higgs self-coupling.

Such a precision measurement of $M_{\ssst{t}}$ or the appearance of new physics between the electroweak and the Planck scale will hopefully 
lead to an answer to the question of vacuum stability in the not too distant future.

\subsection*{Acknowledgements}
I thank K.~G.~Chetyrkin for his collaboration on the project and many invaluable discussions
and J.~H.~K\"uhn for his support and useful comments. 
This work has been supported by the Deutsche Forschungsgemeinschaft in the
Sonderforschungsbereich/Transregio SFB/TR-9 ``Computational Particle
Physics'' and the Graduiertenkolleg 1694 ``Elementarteilchenphysik
bei h\"ochsten Energien und h\"ochster Pr\"azission''.

\bibliographystyle{utphys}
\bibliography{LiteraturSM}

\providecommand{\href}[2]{#2}\begingroup\raggedright\begin{thebibliography}{10}

\bibitem{ATLAS:2012ae}
{\bf ATLAS} Collaboration, G.~Aad {\em et al.}, {\em {Combined search for the
  Standard Model Higgs boson using up to 4.9 fb-1 of pp collision data at
  sqrt(s) = 7 TeV with the ATLAS detector at the LHC}}. Phys.Lett. {\bf B710}
  (2012)  49--66,
\href{http://arxiv.org/abs/1202.1408}{{\tt arXiv:1202.1408 [hep-ex]}}.

\bibitem{Chatrchyan:2012tx}
{\bf CMS} Collaboration, S.~Chatrchyan {\em et al.}, {\em {Combined results of
  searches for the standard model Higgs boson in $pp$ collisions at
  $\sqrt{s}=7$ TeV}}.
  \href{http://dx.doi.org/10.1016/j.physletb.2012.02.064}{Phys.Lett. {\bf B710}
  (2012)  26--48},
\href{http://arxiv.org/abs/1202.1488}{{\tt arXiv:1202.1488 [hep-ex]}}.

\bibitem{pdg2014}
{\bf Particle Data Group} Collaboration, K.~Olive {\em et al.}, {\em {Review of
  Particle Physics}}.
\href{http://dx.doi.org/10.1088/1674-1137/38/9/090001}{Chin.Phys. {\bf C38}
  (2014)  090001}.

\bibitem{PhysRevD.7.1888}
S.~Coleman and E.~Weinberg, {\em Radiative Corrections as the Origin of
  Spontaneous Symmetry Breaking}.
  \href{http://dx.doi.org/10.1103/PhysRevD.7.1888}{Phys. Rev. D {\bf 7} (1973)
  1888--1910}.

\bibitem{Bezrukov:2012sa}
F.~Bezrukov, M.~Y. Kalmykov, B.~A. Kniehl, and M.~Shaposhnikov, {\em {Higgs
  Boson Mass and New Physics}}.
  \href{http://dx.doi.org/10.1007/JHEP10(2012)140}{JHEP {\bf 1210} (2012)
  140},
\href{http://arxiv.org/abs/1205.2893}{{\tt arXiv:1205.2893 [hep-ph]}}.

\bibitem{DiLuzio:2014bua}
L.~Di~Luzio and L.~Mihaila, {\em {On the gauge dependence of the Standard Model
  vacuum instability scale}}.
\href{http://arxiv.org/abs/1404.7450}{{\tt arXiv:1404.7450 [hep-ph]}}.

\bibitem{Altarelli1994141}
G.~Altarelli and G.~Isidori, {\em Lower limit on the Higgs mass in the standard
  model: An update}.
  \href{http://dx.doi.org/10.1016/0370-2693(94)91458-3}{Physics Letters B {\bf
  337} (1994) no.~1-2, 141--144}.

\bibitem{Cabibbo:1979ay}
N.~Cabibbo, L.~Maiani, G.~Parisi, and R.~Petronzio, {\em {Bounds on the
  Fermions and Higgs Boson Masses in Grand Unified Theories}}.
\href{http://dx.doi.org/10.1016/0550-3213(79)90167-6}{Nucl. Phys. {\bf B158}
  (1979)  295--305}.

\bibitem{Ford:1992mv}
C.~Ford, D.~Jones, P.~Stephenson, and M.~Einhorn, {\em {The Effective potential
  and the renormalization group}}.
  \href{http://dx.doi.org/10.1016/0550-3213(93)90206-5}{Nucl. Phys. {\bf B395}
  (1993)  17--34},
\href{http://arxiv.org/abs/hep-lat/9210033}{{\tt arXiv:hep-lat/9210033
  [hep-lat]}}.

\bibitem{Zoller:2014xoa}
M.~Zoller, {\em {Three-loop beta function for the Higgs self-coupling}}. PoS
  {\bf LL2014} (2014)  014,
\href{http://arxiv.org/abs/1407.6608}{{\tt arXiv:1407.6608 [hep-ph]}}.

\bibitem{Zoller:2013mra}
M.~Zoller, {\em {Beta-function for the Higgs self-interaction in the Standard
  Model at three-loop level}}. PoS {\bf (EPS-HEP 2013)} (2013)  322,
\href{http://arxiv.org/abs/1311.5085}{{\tt arXiv:1311.5085 [hep-ph]}}.

\bibitem{Zoller:2012cv}
M.~Zoller, {\em {Vacuum stability in the SM and the three-loop $\beta$-function
  for the Higgs self-interaction}}.
\href{http://arxiv.org/abs/1209.5609}{{\tt arXiv:1209.5609 [hep-ph]}}.

\bibitem{Chetyrkin:2012rz}
K.~Chetyrkin and M.~Zoller, {\em {Three-loop $\beta$-functions for top-Yukawa
  and the Higgs self-interaction in the Standard Model}}.
  \href{http://dx.doi.org/10.1007/JHEP06(2012)033}{JHEP {\bf 1206} (2012)
  033}, \href{http://arxiv.org/abs/1205.2892}{{\tt arXiv:1205.2892 [hep-ph]}}.

\bibitem{Buttazzo:2013uya}
D.~Buttazzo, G.~Degrassi, P.~P. Giardino, G.~F. Giudice, F.~Sala, {\em et al.},
  {\em {Investigating the near-criticality of the Higgs boson}}.
\href{http://arxiv.org/abs/1307.3536}{{\tt arXiv:1307.3536 [hep-ph]}}.

\bibitem{Masina:2012tz}
I.~Masina, {\em {Higgs boson and top quark masses as tests of electroweak
  vacuum stability}}.
  \href{http://dx.doi.org/10.1103/PhysRevD.87.053001}{Phys.Rev. {\bf D87}
  (2013) no.~5, 053001},
\href{http://arxiv.org/abs/1209.0393}{{\tt arXiv:1209.0393 [hep-ph]}}.

\bibitem{Degrassi:2012ry}
G.~Degrassi, S.~Di~Vita, J.~Elias-Miro, J.~R. Espinosa, G.~F. Giudice, {\em et
  al.}, {\em {Higgs mass and vacuum stability in the Standard Model at NNLO}}.
  \href{http://dx.doi.org/10.1007/JHEP08(2012)098}{JHEP {\bf 1208} (2012)
  098},
\href{http://arxiv.org/abs/1205.6497}{{\tt arXiv:1205.6497 [hep-ph]}}.

\bibitem{EliasMiro:2011aa}
J.~Elias-Miro, J.~R. Espinosa, G.~F. Giudice, G.~Isidori, A.~Riotto, {\em et
  al.}, {\em {Higgs mass implications on the stability of the electroweak
  vacuum}}. \href{http://dx.doi.org/10.1016/j.physletb.2012.02.013}{Phys. Lett.
  {\bf B709} (2012)  222--228},
\href{http://arxiv.org/abs/1112.3022}{{\tt arXiv:1112.3022 [hep-ph]}}.

\bibitem{Holthausen:2011aa}
M.~Holthausen, K.~S. Lim, and M.~Lindner, {\em {Planck scale Boundary
  Conditions and the Higgs Mass}}.
  \href{http://dx.doi.org/10.1007/JHEP02(2012)037}{JHEP {\bf 1202} (2012)
  037},
\href{http://arxiv.org/abs/1112.2415}{{\tt arXiv:1112.2415 [hep-ph]}}.

\bibitem{Bobrowski:2014dla}
M.~Bobrowski, G.~Chalons, W.~G. Hollik, and U.~Nierste, {\em {Vacuum stability
  of the effective Higgs potential in the Minimal Supersymmetric Standard
  Model}}. \href{http://dx.doi.org/10.1103/PhysRevD.90.035025}{Phys.Rev. {\bf
  D90} (2014)  035025},
\href{http://arxiv.org/abs/1407.2814}{{\tt arXiv:1407.2814 [hep-ph]}}.

\bibitem{Chetyrkin:2013wya}
K.~Chetyrkin and M.~Zoller, {\em {$\beta$-function for the Higgs
  self-interaction in the Standard Model at three-loop level}}.
  \href{http://dx.doi.org/10.1007/JHEP04(2013)091,
  10.1007/JHEP09(2013)155}{JHEP {\bf 1304} (2013)  091},
\href{http://arxiv.org/abs/1303.2890}{{\tt arXiv:1303.2890 [hep-ph]}}.

\bibitem{Bednyakov:2013eba}
A.~Bednyakov, A.~Pikelner, and V.~Velizhanin, {\em {Higgs self-coupling
  beta-function in the Standard Model at three loops}}.
  \href{http://dx.doi.org/10.1016/j.nuclphysb.2013.07.015}{Nucl.Phys. {\bf
  B875} (2013)  552--565},
\href{http://arxiv.org/abs/1303.4364}{{\tt arXiv:1303.4364}}.

\bibitem{Bednyakov:2013cpa}
A.~Bednyakov, A.~Pikelner, and V.~Velizhanin, {\em {Three-loop Higgs
  self-coupling beta-function in the Standard Model with complex Yukawa
  matrices}}.
\href{http://arxiv.org/abs/1310.3806}{{\tt arXiv:1310.3806 [hep-ph]}}.

\bibitem{PhysRevLett.108.151602}
L.~N. Mihaila, J.~Salomon, and M.~Steinhauser, {\em {Gauge coupling beta
  functions in the standard model to three loops}}.
  \href{http://dx.doi.org/10.1103/PhysRevLett.108.151602}{Phys. Rev. Lett. {\bf
  108} (2012)  151602}.

\bibitem{Mihaila:2012pz}
L.~N. Mihaila, J.~Salomon, and M.~Steinhauser, {\em {Renormalization constants
  and beta functions for the gauge couplings of the Standard Model to
  three-loop order}}. \href{http://dx.doi.org/10.1103/PhysRevD.86.096008}{Phys.
  Rev. D {\bf 86} (2012)  096008}, \href{http://arxiv.org/abs/1208.3357}{{\tt
  arXiv:1208.3357 [hep-ph]}}.

\bibitem{Bednyakov:2012rb}
A.~Bednyakov, A.~Pikelner, and V.~Velizhanin, {\em {Anomalous dimensions of
  gauge fields and gauge coupling beta-functions in the Standard Model at three
  loops}}. \href{http://dx.doi.org/10.1007/JHEP01(2013)017}{JHEP {\bf 1301}
  (2013)  017},
\href{http://arxiv.org/abs/1210.6873}{{\tt arXiv:1210.6873 [hep-ph]}}.

\bibitem{Bednyakov:2012en}
A.~Bednyakov, A.~Pikelner, and V.~Velizhanin, {\em {Yukawa coupling
  beta-functions in the Standard Model at three loops}}.
  \href{http://dx.doi.org/10.1016/j.physletb.2013.04.038}{Phys.Lett. {\bf B722}
  (2013)  336--340},
\href{http://arxiv.org/abs/1212.6829}{{\tt arXiv:1212.6829}}.

\bibitem{Jegerlehner:2012kn}
F.~Jegerlehner, M.~Y. Kalmykov, and B.~A. Kniehl, {\em {On the difference
  between the pole and the MSbar masses of the top quark at the electroweak
  scale}}. \href{http://dx.doi.org/10.1016/j.physletb.2013.04.012}{Phys.Lett.
  {\bf B722} (2013)  123--129},
\href{http://arxiv.org/abs/1212.4319}{{\tt arXiv:1212.4319 [hep-ph]}}.

\bibitem{Espinosa:2007qp}
J.~Espinosa, G.~Giudice, and A.~Riotto, {\em {Cosmological implications of the
  Higgs mass measurement}}.
  \href{http://dx.doi.org/10.1088/1475-7516/2008/05/002}{JCAP {\bf 0805} (2008)
   002},
\href{http://arxiv.org/abs/0710.2484}{{\tt arXiv:0710.2484 [hep-ph]}}.

\bibitem{Hempfling:1994ar}
R.~Hempfling and B.~A. Kniehl, {\em {On the relation between the fermion pole
  mass and MS Yukawa coupling in the standard model}}.
  \href{http://dx.doi.org/10.1103/PhysRevD.51.1386}{Phys.Rev. {\bf D51} (1995)
  1386--1394},
\href{http://arxiv.org/abs/hep-ph/9408313}{{\tt arXiv:hep-ph/9408313
  [hep-ph]}}.

\bibitem{Sirlin1986389}
A.~Sirlin and R.~Zucchini, {\em Dependence of the Higgs coupling hMS(M) on mH
  and the possible onset of new physics}.
  \href{http://dx.doi.org/10.1016/0550-3213(86)90096-9}{Nucl. Phys. B {\bf 266}
  (1986) no.~2, 389--409}.

\bibitem{ATLAS:2014wva}
ATLAS, CDF, CMS, and D0, {\em {First combination of Tevatron and LHC
  measurements of the top-quark mass}}.
\href{http://arxiv.org/abs/1403.4427}{{\tt arXiv:1403.4427 [hep-ex]}}.

\bibitem{Martinez:2002st}
M.~Martinez and R.~Miquel, {\em {Multiparameter fits to the t anti-t threshold
  observables at a future e+ e- linear collider}}.
  \href{http://dx.doi.org/10.1140/epjc/s2002-01094-1}{Eur.Phys.J. {\bf C27}
  (2003)  49--55},
\href{http://arxiv.org/abs/hep-ph/0207315}{{\tt arXiv:hep-ph/0207315
  [hep-ph]}}.

\bibitem{Seidel:2013sqa}
K.~Seidel, F.~Simon, M.~Tesar, and S.~Poss, {\em {Top quark mass measurements
  at and above threshold at CLIC}}.
  \href{http://dx.doi.org/10.1140/epjc/s10052-013-2530-7}{Eur.Phys.J. {\bf C73}
  (2013)  2530},
\href{http://arxiv.org/abs/1303.3758}{{\tt arXiv:1303.3758 [hep-ex]}}.

\end{thebibliography}\endgroup

\end{document}